\documentclass[manuscript]{aastex}

\received{}
\revised{}
\accepted{}

\shorttitle{LMC star field population}
\shortauthors{Piatti et al.}

\begin{document}

\title{The age-metallicity relationship
of the Large Magellanic Cloud field star population from wide-field Washington photometry}

\author{Andr\'es E. Piatti}
\affil{Instituto de Astronom\'{\i}a y F\'{\i}sica del Espacio, CC 67, Suc. 
28, 1428, Ciudad de Buenos Aires, Argentina}
\email{e-mail: andres@iafe.uba.ar}

\and

\author{Doug Geisler}

\affil{Universidad de Concepci\'on, Casilla 160-C, Concepci\'on, Chile}

\begin{abstract}

We analyze age and metallicity estimates for an unprecedented database of some 5.5 million
stars distributed throughout the Large Magellanic Cloud (LMC) main body, obtained from 
CCD Washington $CT_1$ photometry, reported on in Piatti et al. 2012. We produce a comprehensive field star Age-Metallicity 
Relationship
(AMR) from the earliest epoch until 
$\sim$ 1 Gyr ago.
This AMR reveals that the LMC has not evolved chemically  as either a closed-box or bursting system, 
exclusively, but as a combination of both scenarios that have varied in relative strength over the lifetime of the galaxy, although the bursting model falls closer to the data in general.
Furthermore, while old and metal-poor field stars have 
been preferentially formed in the outer disk,  younger and more metal-rich stars have mostly 
been formed in the inner disk, confirming an outside-in formation. We provide  evidence for the formation 
of stars between 5 and 12 Gyr, during the cluster age gap, although chemical enrichment during this
period was minimal. 
We find no significant metallicity gradient in the LMC. We also find that the range in the metallicity 
of an LMC field  
has varied during the lifetime of the LMC. In particular, we find only a small range     of the metal abundance in the outer disk fields, whereas an average range of $\Delta$[Fe/H] = +0.3 $\pm$ 0.1 dex
appears  in the inner disk fields.
Finally, the cluster and field AMRs show a satisfactory 
match only for the last 3 Gyr,  while for the oldest ages ($>$ 11 Gyr) the cluster AMR is a remarkable lower envelope to the 
field AMR. Such a difference may be due to the very rapid early chemical evolution and lack of observed field stars in this 
regime, whereas the globular clusters are easily studied.
This large difference is not easy to explain as coming from
stripped ancient Small Magellanic Cloud (SMC) clusters, although the field SMC AMR is on average $\sim$ 0.4 dex more 
metal-poor at all ages than 
that of the LMC but otherwise very similar.

\end{abstract}


\section{Introduction}

The Large Magellanic Cloud (LMC) has long been recognized as a fundamental
benchmark for a wide variety of astrophysical studies. As the closest bulge-less dwarf disk galaxy
\citep{b12}, it has turned out to be the ideal local
analog for the detailed study of these most common and primeval galaxies. 
Ages and abundances of LMC field star populations are prime indicators of the galaxy's 
chemical evolution and star formation history (SFH). This becomes even more relevant since
its formation and chemical evolution cannot be fully traced from its star cluster populations, due 
to the well-known extended age gap. The LMC age-metallicity
relationship (AMR) has been the subject of a number of studies 
\cite[among others]{oetal91,hetal99,cetal05,retal11}. 
Among them, two perhaps best summarize our current knowledge in this field. First, 
 \citet[hereafter CGAH; see 
also references therein]{cetal11} have examined the AMR for field star populations, based on Calcium
triplet spectroscopy of individual red giants and BVRI photometry in ten 34$\arcmin$$\times$33$\arcmin$ 
LMC fields. They found that:
i) the AMRs for their fields are statistically indistinguishable; ii) the disk AMR is similar 
to that of the LMC star clusters and is well reproduced by closed-box models or models with a small degree of
outflow; iii) the lack of clusters with ages between 3 and 10 Gyr is not observed in the field
population; iv) the age of the youngest population observed in each field increases with galactocentric
distance; v) the rapid chemical enrichment observed in the last few Gyrs is only observed in fields
with R$<$7kpc; vi) the metallicity gradient observed in the outer disk can be explained by an increase
in the age of the youngest stars and a concomitant decrease in their metallicity; and vii) they find much
better evidence for an outside-in than inside-out formation scenario, in contradiction to generic $\Lambda CDM$ models.
Secondly,  \citet[hereafter HZ09; see 
also references therein)]{hz09} presented the first-ever global,
spatially-resolved reconstruction of the SFH, based on the application of their
StarFISH analysis software to the multiband photometry of twenty million stars from the Magellanic
Clouds Photometric Survey. They found that there existed a long relatively quiescent epoch (from $\sim$ 12 to 5 Gyr ago) 
during which the star formation was suppressed throughout the LMC; the metallicity also remained stagnant during 
this period. They
concluded that the field and cluster star formation modes have been tightly coupled throughout the LMC's history.

Although these studies represent the state-of-the-art of our knowledge of the  LMC AMR, they leave
unanswered a number of outstanding questions:
What caused the general lull in  SF between $\sim$ 5 and 12 Gyr ago? Are the cluster and field AMRs really tightly 
coupled? Can the LMC AMR best be described by a closed-box, bursting or other chemical evolution model? What, if any, 
are the radial dependences?
In addition, HZ09 did not go deep enough to derive the full SFH from information on the Main Sequence 
(MS). They reached a limiting magnitude between $V$ = 20 and  21 mag, depending on the local degree of 
crowding in the images, corresponding to stars younger than 3 Gyr old on the MS if the theoretical isochrones 
of \citet{getal02} and a LMC distance modulus of 18.5 mag are used. 
Thus, the
advantages of covering an enormous extension of the LMC is partially offset by the loss in depth of the
limiting magnitude. On the other hand, the ten fields of CGAH cover a rather small fraction of the whole
LMC. Therefore, it is desirable to obtain an overall deeper AMR for the LMC which also covers a larger area. Previous 
AMRs have been founded on 
theoretical isochrones, numerical models, or synthetic Color-Magnitude Diagrams (CMDs), so that an AMR 
built from actual measured ages and metallicities is very valuable. A comprehensive comparison between the
field and cluster AMRs obtained using the same procedure is also lacking. All these aims demand
the availability of a huge volume of high quality data as well as a powerful technique to provide both
accurate ages and metallicities.

In this paper we address these issues for the first time. We make use of an unprecedented 
database of some 5.5 million stars measured with the Washington $CT_1$ photometric system, which are 
spread over a large part of the LMC main body. From this database, we produce the LMC field AMR from the birth of 
the galaxy until $\sim$ 1 Gyr ago, using the $\delta$$T_1$ index and the standard  giant branch isoabundance
curves to estimate ages and metallicities, respectively, of the most representative field populations.
These provide approximately independent measurements of these two quantities, minimizing the
age-metallicity degeneracy problem. 
In addition, this is the 
first overall LMC field star AMR obtained from Washington data; thus complementing those derived from other data sets
such as HZ09 or the AMR obtained from Washington data for LMC clusters \citep{p11a}.
Finally, we homogeneously compared the derived field star AMR to that for the LMC cluster
population with ages and metallicities put on the same scales using these two Washington datasets. This kind of 
comparison has not been accomplished 
before. The paper is organized as follows: Section 2 briefly describes the data handling and analysis from which 
\citet{pgm12} estimated  the field star ages and metallicities. Section 3 deals with the aforementioned issue of 
a comprehensive AMR of the LMC field star population. In Section 4 we discuss our results and compare them with
previous studies, while Section 5 summarizes our major findings.

\section{Data handling and scope}

We obtained Washington photometric data
at the Cerro-Tololo Inter-American Observatory 
(CTIO) 4 m Blanco telescope with the Mosaic \,II camera attached (36$\arcmin$$\times$36$\arcmin$ field onto a
8K$\times$8K CCD detector array) of twenty-one LMC fields, concentrated in the main body but mostly
avoiding the very crowded bar regions. 
We refer the reader to  \citet{pgm12} for details about the observations and 
reduction and analysis of the data. Briefly, Piatti et al. analysed the $C$ and $T_1$ limiting magnitudes reached for 
a 50\% completeness level from extensive artificial star tests, 
produced CMDs, Hess-diagrams, MS star luminosity functions, Red Clump star $T_1$ mag histograms, 
RGB distributions, etc, and presented a thorough description of the uncertainties involved and of the techniques used.
 The processed data are much deeper than those used by HZ09 and generally 
reach well below the MS Turnoffs (TOs) of the oldest stellar populations in the LMC 
($T_{1_o}$ $\sim$ 19.9 - 21.4 mag). In addition, the
total area covered is about 2.5 times larger that that of CGAH.
We subdivided each 36$\arcmin$$\times$36$\arcmin$ field into 16 uniform 2K$\times$2K regions 
(9$\arcmin$$\times$9$\arcmin$ each).

The stellar photometry was performed using the 
star-finding and point-spread-function (PSF) fitting routines in the {\sc DAOPHOT/ALLSTAR} suite of programs 
\citep{setal90}. 
The standard Allstar - Find - Subtract procedure was repeated three times for each frame. 
Finally, we combined all the independent measurements of the stars in the different filters
 using the stand-alone {\sc DAOMATCH} and {\sc DAOMASTER}
programmes, kindly provided by Peter Stetson.

 \citet{pgm12} used the so-called "representative" population, defined in 
\citet{getal03}, to measure ages for the 21 fields in the same way as \citet{p12} did 
for 11 fields of the Small Magellanic Cloud (SMC). 
\citet{getal03} assumed that the observed MS in each LMC field 
is the result of the superposition of MSs with different TOs (ages) and 
constant luminosity functions. Thus, the intrinsic number of stars belonging to any MS interval comes from
the difference of the total number of stars in that interval and that of the adjacent intervals. 
Therefore, the biggest difference is directly related to the most populated TO. This "representative" AMR differs 
from those derived from modeled SFHs in the fact 
that it does not include complete information on all stellar populations, but accounts for the dominant population 
present in each field. Minority populations  are not considered, nor dominant populations younger than $\sim$ 1 Gyr, due to our inability to age-date them.
The method has turned out to be a powerful tool for revealing the primary trends in an efficient and robust way 
\citep{petal03a,petal03b,petal07}.

\citet{pgm12} clearly identified the representative star populations in the 21 studied LMC fields, 
which were typically $\sim$ 25\%-50\% more frequent than the second most numerous population. They derived ages from 
the $\delta$$T_1$ index, calculated by determining the difference in the $T_1$ magnitude of the red clump (RC) and 
the representative MS TO \citep{getal97}.  The $\delta (T_1)$ index has proven to be a powerful tool to derive 
ages for star clusters older than 1 Gyr, independently of their metallicities \citep{betal98,petal02,petal09,petal11a,p11a}. 
Indeed,  \citet{getal97} showed that $\delta$($T_1$) is very well-correlated with $\delta$($R$) (correlation coefficient 
= 0.993) and with $\delta$($V)$. We then derived ages from the 
$\delta T_1$ values using equation (4) of  \citet{getal97}, which was obtaining by fitting $\delta (T_1)$ values of
star clusters with well-known age estimates. This equation is only calibrated
for ages larger than 1 Gyr, in particular because the magnitude of the He-burning  stage varies with age
for such massive stars, so that we are not able to produce ages for younger representative populations. Note that 
this age measurement technique does not require 
absolute photometry and is independent of reddening and distance as well. An additional advantage is that we do not 
need to go deep enough to see the extended MS of 
the representative star population but only slightly beyond its MS TO. The representative MS TO $T_1$ magnitude for 
each subfield turned out to be 
on average $\sim$ 0.6 mag brighter than the $T_1$ mag for the 100$\%$ completeness level of the respective subfield,
so that Piatti et al. actually 
reach the TO of the representative population of each subfield. 
Note that the representative stellar population is not
necessarily the oldest one reached in a subfield. Their Figs. 3 to 23 illustrate
the performance of their photometric data. 
In their Table 5, they presented
the final ages and their dispersions. Such dispersions have been calculated bearing in mind the
broadness of the $T_1$ mag distributions of the representative MSTOs and RCs, and not just simply the photometric errors
at $T_1$(MSTO) and $T_1$(RC) mags, respectively. The former are clearly larger
and represent in general a satisfactory estimate of the age spread around the prevailing population ages, although 
a few individual subfields have a slightly larger age spread.  We refer the reader to the companion paper by \citet{pgm12}
for details concerning the methods and limitations and uncertainties involved.

The mean metallicity for each representative field population was 
obtained by first entering the positions of the representative giant branch 
into the [$M_{T_1}$, $(C-T_1)_o$] plane with the Standard Giant Branch (SGB) isoabundance curves traced by 
\citet{gs99}. This was done to obtain, by interpolation, metal abundances
 ([Fe/H]) with typical errors of $\sim$ 0.20 dex. Then, they applied the appropriate age correction
to these metallicities using the age-correction procedure of  \citet{getal03}, which provides 
age-corrected metallicities in good general agreement with spectroscopic values \citep{petal10}.
The resulting metallicities and their dispersions are compiled in Table 6 of \citet{pgm12}.
Tables 5 and 6 of Piatti et al. 2012 are reproduced here as Tables 1 and 2 for completeness sake.

\section{The AMR}

One of the unavoidable complications in analysing measured ages and metallicities is that they have associated uncertainties. 
Indeed, by considering such errors, the interpretation
of the resulting AMR can differ appreciably from that obtained using only the measured ages and metallicities without
 accounting for their errors. However, the treatment of
age and metallicity errors in the AMR is not a straightforward task. Moreover, even if errors did not play an 
important role, the binning of age/metallicity
ranges could also bias the results. For example, fixed age intervals have commonly been used to build cluster age 
distributions using the same cluster database \citep{botal06,wetal09,p10}, with remarkably different results depending on 
the details of the binning process.
These examples show that a fixed age bin size is not appropriate for yielding the intrinsic age distribution, since 
the result depends on the chosen age interval and the age errors are typically a strong function of the  
age. A more robust age bin is
one  whose width is of the order of the age errors of the clusters in that interval.
This would lead to the selection of very narrow bins (in linear age) for young clusters and relatively broader age
bins for the older ones.

With the aim of building an age histogram that best reproduces the intrinsic open cluster
age distribution, \citet{p10} took the uncertainties in the age estimates into account in order
to define the age 
intervals in the whole Galactic open cluster age range. Thus, he produced a more appropriate sampling of the open clusters
 per age interval than is generated using a fixed bin size, since he included in each bin a number of clusters whose age 
errors are close to the size of this bin. Indeed, the age errors for very young clusters are 
 a couple of Myrs,
 while those for the oldest clusters are at least a few Gyrs. Therefore, smaller bins are appropriate for 
young clusters,
whereas larger bins are more suitable for the old clusters.  \citet{petal11a,petal11b}
have also used this precept for producing age distributions of LMC and SMC clusters, respectively.

We then searched Table 5 of \citet{pgm12} (the present Table 1) to find that 
typical age errors are  0.10 $\la$ $\Delta$log($t$) $\la$ 0.15. Therefore, we produced the AMR of the LMC
field population by setting the age bin sizes according to this logarithmic law, which
traces the variation in the derived age uncertainties in terms of the measured ages. We used 
intervals of $\Delta$log($t$) = 0.10. We proceeded in a similar way when binning
the metallicity range. In this case, we adopted a [Fe/H] interval of 0.25 dex. Thus, the subdivision of the 
whole age and metallicity ranges was then performed on an observational-based foundation, 
since the (age,[Fe/H]) dimensions are determined by the typical errors for each age/metallicity range. 
However, there is still an additional issue to be
considered: even though the (age,[Fe/H]) bins are set to match the age/metallicity errors,  
any individual point in the 
AMR plane may fall in the respective (age,[Fe/H]) bin or in any of the eight adjacent bins. This happens
when an (age,[FeH]) point does not fall in the bin centre and, due to its errors,  has 
the chance to fall outside it. Note that, since we chose bin dimensions as large as the involved errors, such points 
should not fall on average far beyond the adjacent bins. However, this does not necessarily happen to all 336 (age,[FeH]) points, and  we should consider at the same time any other possibility. 

We have  taken all these effects into account to produce the AMR of the 21 studied LMC fields. 
First of all, we take the AMR plane as engraved by a grid of bins as mentioned above, i.e. with logarithmic and linear
scales drawn along the age and metallicity axes. Then, if we put one of our (age,[Fe/H]) points in it, we find out that
that point with its errors covers an area which could be represented by a box of size 
4$\times$$\sigma(age)$$\times$$\sigma([Fe/H])$. This (age,[Fe/H]) box may or may not fall centered on one of the AMR grid 
bins, and has dimensions smaller, similar or larger than the AMR grid bin wherein it is placed. Each
of these scenarios generates a variety of possibilities, in the sense that the (age,[Fe/H]) box could cover from  
one up to 25 or more AMR bins depending on its position and size. Bearing in mind all these alternatives, our strategy 
consisted in weighing the contribution of each (age,[Fe/H]) box to each one of the AMR grid bins occupied by it, so that
the sum of all the weights equals unity. 
The assigned weight was computed as the ratio between the area occupied by the (age,[Fe/H]) box in a AMR grid bin to the 
(age,[Fe/H]) box size. When performing such a weighting process, we focused in practice on a single AMR grid bin and
calculated the weighted contribution of all the 336 (age,[Fe/H]) boxes to that AMR grid bin. We then repeated the calculation 
for all the AMR grid bins. In order to know whether a portion of an (age,[Fe/H]) box falls in a AMR grid bin, we
took into account the following possibilities and combinations between them. Once an age interval is defined, we asked
whether: i) the age associated with any of the 336 (age,[Fe/H]) points is inside that age interval, ii) the age-$\sigma$(age)
value is inside that age interval; iii) the age-$\sigma$(age) value is to the left of that age interval and the age is
to the right; iv) the age+$\sigma$(age) value is inside that age interval and, v) the age+$\sigma$(age) value is to the right
 of that age interval and the age is to the left. For the metallicities we proceeded in a similar way so that we finally
encompassed a total of 25 different inquiries to exactly match the positions and sizes of the 336 (age,[Fe/H]) points in the
AMR plane grid. We are confident that our analysis yields accurate morphology
and position of the main features in the derived AMRs. 

Fig. 1 shows the resulting individual AMRs as labelled at the top-right margin of each panel. 
It is important to keep in mind that each of the (age,[Fe/H]) points used to make each of these plots is 
simply the representative, most dominant population in that subfield.
The filled boxes
represent the obtained mean values for each (age,[Fe/H]) bin; the age error bars follow the law $\sigma$log($t$) = 0.10;
 and the [Fe/H]
error bars come from the full width at half-maxima (FWHMs) we derived by fitting Gaussian functions to the metallicity 
distribution in each age interval.
The fit of a single Gaussian per age bin was performed using the NGAUSSFIT routine in the STSDAS/IRAF\footnote{IRAF is 
distributed by the National Optical Astronomy Observatories, 
which is operated by the Association of Universities for Research in Astronomy, Inc., under contract with 
the National Science Foundation} package. The centre 
of the Gaussian, its amplitude and its FWHM acted as variables, while the constant and the linear terms were fixed to zero, 
respectively. We used Gaussian fits for simplicity. We estimated a difference from Gaussian distributions of only $\approx$
8 $\%$.
At first glance, it can be seen that the youngest and the oldest ages of each AMR vary from field to field.
The metallicity range and  the shapes of the 21 AMRs are also quite variable. For example, AMRs for Fields $\#$
3, 6, and 8 do not show chemical enrichment, a feature that can be seen for example in Fields $\#$ 10, 12, 13, and 
14. Moderate to intermediate chemical enrichment is seen in the remaining fields. Fields $\#$14 and 20 are
the most metal-rich and the most metal-poor fields, respectively, at any time, with a  mean difference between them of $\sim$
 0.8 dex. 

In order to examine whether there exists any dependence of the individual AMRs with position in the LMC, 
we have made use of their deprojected galactocentric distances computed by assuming that they are part of 
a disk having an inclination $i$ = 35.8$\degr$ and a position angle of the line 
of nodes of $\Theta$ = 145$\degr$ \citep{os02}. We refer the reader to Table 1 of \citet{ss10} which includes a summary of 
orientation measurements of the LMC disk plane, as well as their analysis of the orientation and other LMC disk quantities, supporting 
the present adopted values. Figs. 2 and 3 illustrate 
the behaviour of the old and the young extremes and the
metal-poor and the metal-rich extremes of each AMR, respectively, as a function of the deprojected distance. Old and metal-poor 
extremes are drawn with open boxes, while young and metal-rich extremes are depicted with filled boxes. The error bars for ages and
metallicities are those from Fig. 1, whereas the error bars of the deprojected distances come from the dispersion of this 
quantity within the 16 subfields used in each mosaic field. As can be seen, the outer fields -defined as those with
deprojected distances $>$ 4$\degr$ \citep[][and references therein]{betal98}- contain dominant stellar populations about as old as the galaxy, while those of the inner
disk do not, with the exception of Fields $\#$9 and 18. The outer fields began at an age within our oldest age interval,
although we have represented them as a single value as a result of our binning process.
In general, the oldest dominant stellar populations in the inner disk
fields have been formed between $\sim$ 5 and 8 Gyr ago. Likewise, the main stellar formation processes in the outer disk 
appears to have ceased some 5 $\pm$ 1 Gyr ago. This result confirms that of \citet{getal08} and CGAH concerning
outside-in evolution of the LMC disk as opposed to the $\Lambda CDM$ prediction for inside-out 
formation.
It is interesting that the inner fields appeared to start their first strong star formation episodes at about the same time 
that the outer fields were undergoing their last episode. This is certainly not what is expected
if the impulse driving the onset of star formation is some global effect like a close galactic encounter,
e.g. with the Galaxy or the SMC. The last major epoch of star formation we are sensitive to ended 
about 1-2 Gyrs ago in the inner fields, with evidence for a radial age gradient. 

On the metallicity side, Fig. 3 shows that for outer fields starting (open box) and ending (filled box) [Fe/H] values
are very similar, which means that they have not experienced much chemical enrichment. Taking into account
the open and filled boxes for these outer fields, we derived a mean value of [Fe/H] = 
-0.90 $\pm$ 0.15 dex (note that their mean starting and ending metallicities are [Fe/H] = -0.95$\pm$0.10 dex and 
-0.90$\pm$0.10 dex, respectively). This value could be considered as the representative
 metallicity level for the outer disk field 
during the entire life of the LMC. In the inner disk, the situation is different. Firstly, the starting metal
abundances (open boxes) are on average as metal-poor as the ending abundances for the most metal-rich outer fields. Secondly,
 there exists a mean increase   in the [Fe/H] values of +0.3 $\pm$ 0.1 dex, indicating significant chemical enrichment. If, in 
addition, we
consider that these inner disk fields have been formed more recently than those in the outer disk, the signs of significant 
recent chemical
enrichment are even more evident. 

The apparent metallicity gradient exhibited in Fig. 3, in the sense that
the more distant a field from the galaxy centre, the more metal-poor it is, is
tightly coupled with the relationship shown in Fig. 2. To disentangle both dependences we fit the 336 
individual metallicities (Table 2) according to the
following expression:

\begin{equation}
{\rm [Fe/H]} = C + (\partial{\rm [Fe/H]}/\partial t)\times t + (\partial{\rm [Fe/H]}/\partial a)\times a
\end{equation}
 
\noindent where $t$ and $a$ represent the age in Gyr and the deprojected distance in degrees. 
The respective
coefficients turned out to be $C$ = -0.55 $\pm$ 0.02 dex, $\partial${\rm [Fe/H]}/$\partial$$t$ = -0.047 $\pm$
0.003 dex Gyr$^{-1}$, and $\partial${\rm [Fe/H]}/$\partial$$a$ = -0.007 $\pm$ 0.006 dex degrees$^{-1}$,
which implies a small but insignificant metallicity gradient of (-0.01 $\pm$ 0.01) dex kpc$^{-1}$, if an LMC 
distance of 50 kpc is adopted 
\citep{ss10}. Thus, there is no evidence for a significant  metallicity gradient in the 
LMC. This result agrees with that of \citet{getal06} who found that the LMC lacks any metallicity gradient. 
The relatively more metal-poor stars found in the outermost regions (see
Fig. 3) are mostly  a consequence of the fact that such regions are dominated by old stars which are relatively metal-poor, 
whereas intermediate-age stars which are more metal-rich prevail in the innermost regions. This result 
firmly confirms CGAH's findings. 

We have also produced a composite AMR for the 21 LMC fields following the
same procedure used to derive the individual AMRs of Fig. 1. The result is shown in Fig. 4, where the mean points
are represented with filled boxes, while the error bars are as for Fig. 1. We have also included the individual points
of the 336 subfields plotted with gray-scale colored triangles. We used a 100 level gray-scale from black to white to
represent the most distant to the nearest star fields to the LMC centre. As can be seen, the most distant fields have been
preferentially formed at a low and relatively constant metallicity level, from the birth of the LMC until $\sim$ 6 Gyr 
ago, while the inner fields have been formed later on with a steeper chemical enrichment rate. Note also that
the [Fe/H] errorbars cover a larger range than that the points represent. This is because these errorbars
do not only represent the standard dispersion of the points, but also of their measured errors (see Sect. 3).

\section{Comparison and discussion of the LMC AMR}

In Fig. 5, we have overplotted with solid lines different field star AMRs along with our presently derived composite AMR, 
namely: HZ09 (yellow),  \citet{retal11} (black), 
\citet[hereafter PT98]{pt98} (blue), and  \citet{getal98} (red). 
The red line AMR is based on a closed-box model, while
the blue curve is a bursting model. We also included with red and blue filled circles the AMRs derived by 
\citet{cetal08} for the LMC bar and disk, respectively. At first glance, we find that the bursting SFH modeled by
PT98 appears to be the one which best resembles the AMR derived by \citet{cetal08}, instead of
closed-box models as Carrera et al. suggested. However, such a resemblance is only apparent since PT98 constructed their 
model using nearly no star formation from $\sim$ 12 up to 3 Gyr ago (see their Fig. 2). This clearly contradicts  
not only Carrera et al.'s result but also ours, which show that there were many stars formed in the  
LMC in that period (see Fig. 4). Indeed, we actually see no significant chemical evolution from about 12 - 6 Gyr,
even though stars were formed. In turn, the closed-box models appear to be qualitatively closer to HZ09's reconstructed AMR. 

Since HZ09's AMR is based on a relatively bright limiting magnitude database and CGAH's AMRs rely on ages and metallicities
 for stars distributed in ten fields (each only slightly smaller than ours), we believe that the present composite AMR has
 several important advantages over these
previous ones, and possibly reconciles previous conclusions about the major 
enrichment processes that have dominated the chemical evolution of the LMC from its birth until $\sim$ 1 Gyr 
ago. Note that a large number of fields distributed through the galaxy are analysed here and their representative 
oldest MS TOs are well measured in all fields. The composite AMR we derive results in a complex function having HZ09's AMR (or 
alternatively the closed-box model) and CGAH's AMRs (or alternatively the bursting model) as lower and 
approximately upper envelopes in metallicity, respectively, although the bursting model is a much better
fit. Therefore, we find evidence that the LMC  has not chemically 
evolved as a closed-box or bursting system, exclusively, but as a combination of both scenarios that likely have varied in 
importance during
the lifetime of the galaxy, but with the bursting model dominating. The closed-box model presumably reproduces the metallicity
trend that the LMC would have had if bursting formation episodes had not taken place. However, since the LMC
would appear to have experienced such an enhanced formation event(s), important chemical enrichment has occurred from
non well-mixed gas spread through the LMC. CGAH  also found that the AMRs for their ten fields are statistically 
indistinguishable. We note, however, that six of their fields are aligned somewhat perpendicular to the LMC bar, reaching
 quite low density outer regions, and 
therefore, that their coverage represents a relatively small percentage of the whole field population. We show in Fig. 1
that, when more field stars distributed through the LMC are analyzed with age and metallicity
uncertainties robustly considered, distinct individual AMRs do arise. Indeed, Figs. 2 and 3 illustrate how different AMRs are
for inner and outer fields.

When inspecting in detail our composite field LMC AMR, the  relatively quiescent epoch ($t$ $\sim$ 5 to 12 Gyr) claimed by 
HZ09 and also frequently considered as a feature engraved in the cluster formation processes, i.e.
the cluster age-gap 
\cite[among others]{getal97,petal02,betal04} is not observed. On the contrary, there exists a noticeable number of 
fields with representative ages spanning the age gap (from $\sim$ 12 Gyr to 3 Gyr), which further strengthens the difference between cluster and field star formation during this epoch. Of course, we
do not quantitatively compare the level of SF in different epochs, we simply measure the properties of
the dominant population. 
However, during this extended period, although some star formation occurred, it was 
not accompanied by any significant chemical evolution until starting $\sim$6 Gyr ago. Again curiously, 
there were several Gyr of star formation and chemical evolution before the cluster age gap ended. In 
addition, although the ages estimated by CGAH  of field stars spanning the cluster Age Gap could have uncertainties 
necessarily large for individual stars, and consequently their SFH would still indicate a relatively  quiescent epoch between 
5 and 10 Gyr as HZ09 pointed out, we provide here evidence of the existence of 
stars formed between 5 and 12 Gyr  that  represent the most numerous 
populations in their respective regions. Note that our metallicities are generally about 0.1 - 0.2 dex lower
than CGAH for younger ages but higher for the oldest stars, indicating a smaller total chemical enrichment over the
lifetime of the galaxy compared to that found by CGAH. Our agreement with \citet{retal11} is somewhat
better. We also find that the amount of chemical evolution (as measured
by the increase   in the metallicity) of the LMC fields  
has varied during the lifetime of the LMC. Particularly, we find only a small range     of the metal abundance
within the considered uncertainties for the outer disk fields, whereas an average increase of $\Delta$[Fe/H] = 0.3 $\pm$ 0.1 dex
appears  in the inner disk fields, and this increase occurred over a relatively shorter time period. Hence, a bursting star 
formation scenario turns out to 
be a plausible explanation if the enhanced star formation is accompanied by a vigorous nucleosynthesis 
process that takes place during the burst. 


Finally, we present a homogeneous comparison between the composite field AMR with that for 81 LMC clusters with 
ages ($\ga$ 1 Gyr) and metallicities derived on the same scales as here. We use the ages and metallicities compiled by 
 \citet{petal11a} for 45 clusters observed in the Washington system, to which we add 36 clusters with ages
 estimated by 
\citet{p11c} from similar data. We estimate here 
their metallicities following the same procedure used for the 
studied fields (see Section 2). 
The resulting cluster AMR is depicted in Fig. 6 with dark-gray filled boxes superimposed onto the composite field LMC
(open boxes with error bars). As can be seen, the cluster AMR satisfactorily matches the field AMR only for the last 3 Gyr, 
while it is a remarkable lower envelope of the field AMR  for older ages ($t$ $>$ 11 Gyr). 
 The most likely explanation is a very rapid early chemical enrichment traced by the very visible globular clusters, but their coeval, low metallicity field counterparts are so rare that they are missed in our data. The origin of the 15
oldest LMC clusters still remains unexplained and constitutes one of the most intriguing enigmas in our understanding of the 
LMC formation and evolution. Different studies show that they have very similar properties to 
the globular clusters 
in the Milky Way \citep[][among others]{betal96,mg04,vdbm04,metal09,metal10}, except for their orbits, which are within the 
LMC disk instead of in an isothermal halo \citep{b07}. On the other hand, Fig. 4 show that there exist field star populations 
older than 10 Gyr and about as old as the old globulars. These results go along with the curious conundrum of the absence of clusters 
during the infamous Age Gap \citep{betal04}.
Since HZ09 found that 
there was a  relatively quiescent epoch in the field star formation from approximately 12 to 5 Gyr ago (similar to that observed for
star clusters), they also concluded that field and cluster star 
formation modes are tightly coupled. Notice that the ages and metallicities used by HZ09 for the 85 clusters 
are not themselves on a homogeneous scale nor on the same field age/metallicity scales. 

In order to look for clues for the very low metallicities of the oldest LMC clusters, we reconstructed the cluster and 
field AMRs of the SMC, also from Washington photometry obtained by us. As for the field AMR we used the ages and metallicities
 derived by 
\citet[his Table 4)]{p12} and applied to them the same binning and error analyses as for the composite LMC field 
AMR (Fig. 4). Note that these ages and metallicities are all set on the present age/metallicity scales. 
We also compiled 59 SMC clusters ($t$ $\ge$ 1 Gyr) from \citet{petal11b}, and 
\citet{p11a,p11b}  
with ages and metallicities tied to the same scales. Fig. 6 shows the resulting SMC AMRs depicted
with open triangles
for its field stars and with filled triangles for its star cluster population. As can be seen, cluster and field stars 
apparently share similar chemical enrichment histories in the SMC, although the population of old clusters drastically 
decreases beyond
 $\sim$ 7 Gyr and there is only 1 older than 10 Gyr. \citet{p11b} showed, based on the statistics of
catalogued and studied clusters, that a total of only seven relatively old/old clusters remain to be studied, and an even 
smaller number is obtained if the cluster spatial distribution is
considered. From this result, we conclude that the SMC cluster AMR is relatively well-known, 
particularly towards its older and more metal-poor end. Therefore, it does not seem easy to connect the origin of the 
oldest LMC cluster population to stripping events of ancient SMC star clusters. Moreover, the composite SMC field AMR
is on average $\sim$ 0.4 dex more metal-poor at all ages than that of the counterpart in the LMC, with
little variation, indicating that the global chemical evolution in these two galaxies was quite similar in
nature but with an offset to lower metallicity in the SMC. In particular, there was a very early and rapid
period of enrichment, followed by a long quiescent epoch with some star formation in both Clouds but
cluster formation only in the SMC and little to no metallicity increase
and finally a recent period of substantial
enrichment starting about 6Gyr ago. This is in very good agreement with the SMC AMR found by
\citet{petal10}. 
The relative deficiency in heavy elements of the SMC could explain the metallicity of a few old LMC clusters, if they were captured
from the SMC \citep{betal12}, but this is an unlikely argument to explain the majority of them. In fact,
it is curious in this context that the the oldest SMC cluster is at the young and metal-rich extreme of the LMC globular cluster distributions.

\section{Summary}

In this study we present, for the first time, the AMR of the LMC field star population from ages and metallicities
derived using CCD Washington $CT_1$ photometry of some 5.5 million stars
in twenty-one 36$\arcmin$$\times$36$\arcmin$ fields distributed throughout the LMC main body
presented in \citet{pgm12}.
The analysis of the photometric data -subdivided in 336 smaller 9$\arcmin$$\times$9$\arcmin$ subfields - leads to 
the following main conclusions:

i) From ages and metallicities of the representative star population in each subfield estimated by using the 
$\delta$$T_1$ 
index and the SGB technique, respectively, we produced individual field AMRs with a robust treatment of their age and
metallicity uncertainties.  These individual AMRs show some noticeable differences from field to field in several 
aspects: starting and ending ages, metallicity range    , shape, etc. This is contrary to CGAH, who found
very similar AMRs in their sample.
The composite AMR for the LMC fields
reveals that, while old and metal-poor field stars have been preferentially formed in the outer disk, younger and more 
metal-rich stars have mostly been formed in the inner disk. This result confirms an outside-in
evolution of the galaxy, as found by \citet{cetal08}. In addition, we provide  evidence of the existence 
of stars formed between 6 and 12 Gyr that represent 
the most numerous populations in their respective regions, although little or no chemical evolution occured during this 
extended period.

ii) The resulting distribution of the ages and the metallicities as a function of the deprojected distance reveals
that there is no significant metallicity gradient in the LMC ((-0.01 $\pm$ 0.01) dex kpc$^{-1}$).
The relatively more metal-poor stars found in 
the outermost regions is mainly a consequence of the fact that such regions are dominated by old stars which 
are relatively 
metal-poor, 
whereas intermediate-age stars which are more metal-rich prevail in the innermost regions. 
We also find that the range     in the metallicity of the LMC fields  
has varied during the lifetime of the LMC. In particular, we find only a small range     of the metal abundance for the outer disk fields, whereas an average range of 
$\Delta$[Fe/H] = +0.3 $\pm$ 0.1 dex is found in the inner disk fields.

iii) From the comparison of our composite AMR with theoretical ones, we conclude that the LMC has not chemically 
evolved as a closed-box or bursting system, exclusively, but as a combination of both scenarios that have
had different prominence during the lifetime of the galaxy, with the bursting model generally more
dominant.  Enhanced formation episodes could have possibly taken place
as a result of its interactions with the Milky Way and/or SMC.

iv) We finally accomplish a homogeneous comparison between the composite field AMR with that for LMC clusters with ages
and metallicities on the same scales. We find a satisfactory match only for the last 3 Gyr, while for older
ages ($>$ 11 Gyr) the cluster AMR results in a remarkable lower envelope of the field AMR. 
The most likely explanation is a very rapid early chemical enrichment traced by the very visible globular
clusters, but their coeval, low metallicity field counterparts are so rare that they are missed in our data.
We find that such a
large difference between the metallicities of LMC field stars and clusters is not easy to explain as coming from
stripped ancient SMC clusters, although the field SMC AMR is on average $\sim$ 0.4 dex more metal-poor at all ages than 
that of the LMC. The two galaxies otherwise show a very similar chemical evolution.

\acknowledgements
We greatly appreciate the comments and suggestions raised by the
reviewer which helped us to improve the manuscript.
This work was partially supported by the Argentinian institutions CONICET and
Agencia Nacional de Promoci\'on Cient\'{\i}fica y Tecnol\'ogica (ANPCyT). 
D.G. gratefully acknowledges support from the Chilean 
BASAL   Centro de Excelencia en Astrof\'{\i}sica
y Tecnolog\'{\i}as Afines (CATA) grant PFB-06/2007.

\begin{figure}
\includegraphics[angle=0,width=17cm]{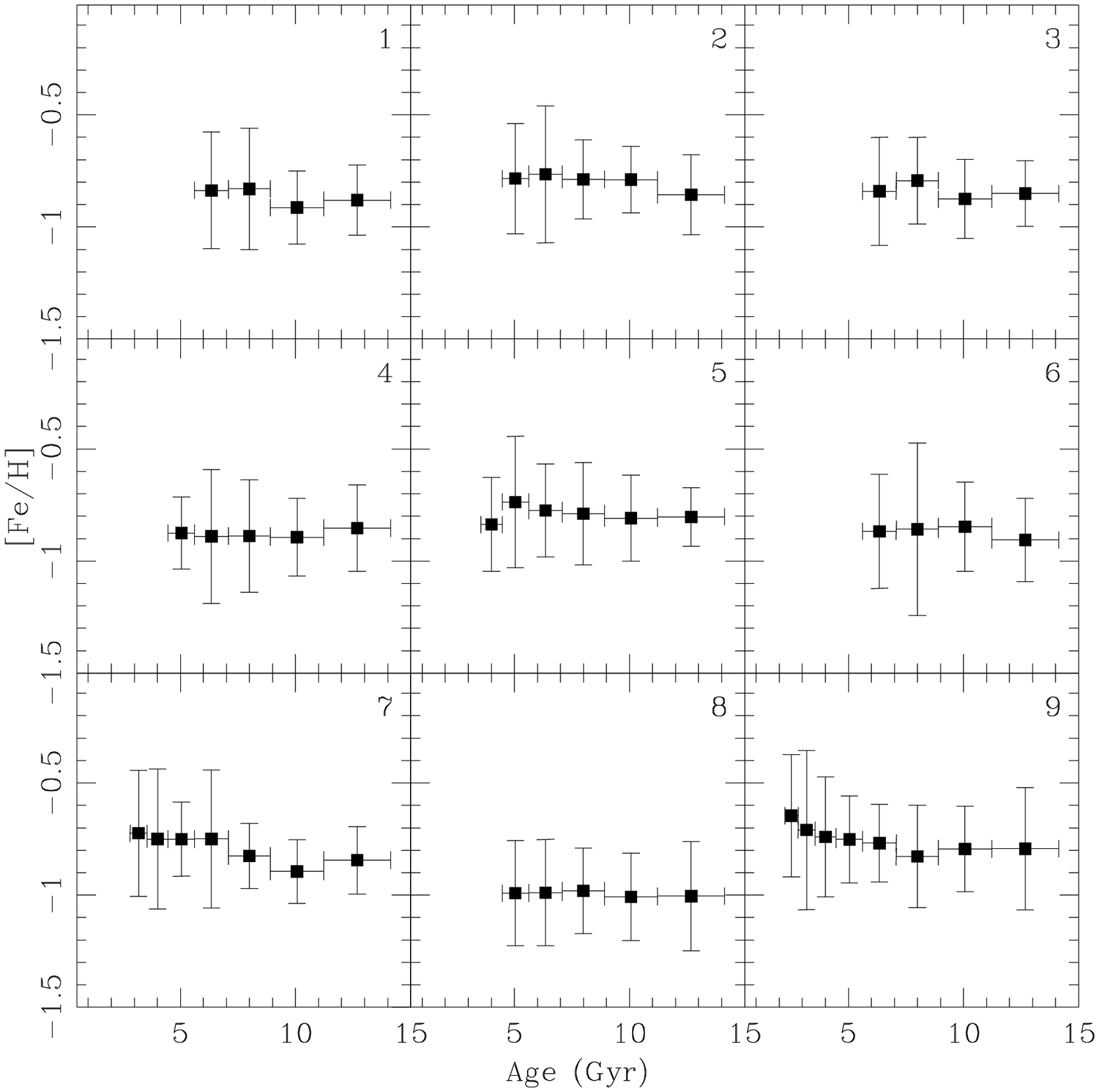}
\caption{The resulting AMRs for the 21 studied LMC fields. The age error bars 
follow the law $\sigma$log($t$) = 0.10, while the [Fe/H]
error bars come from the FWHMs of Gaussian functions fitted to the metallicity 
distribution in each age interval.}
\label{fig1a}
\end{figure}

\setcounter{figure}{0}
\begin{figure}
\includegraphics[angle=0,width=17cm]{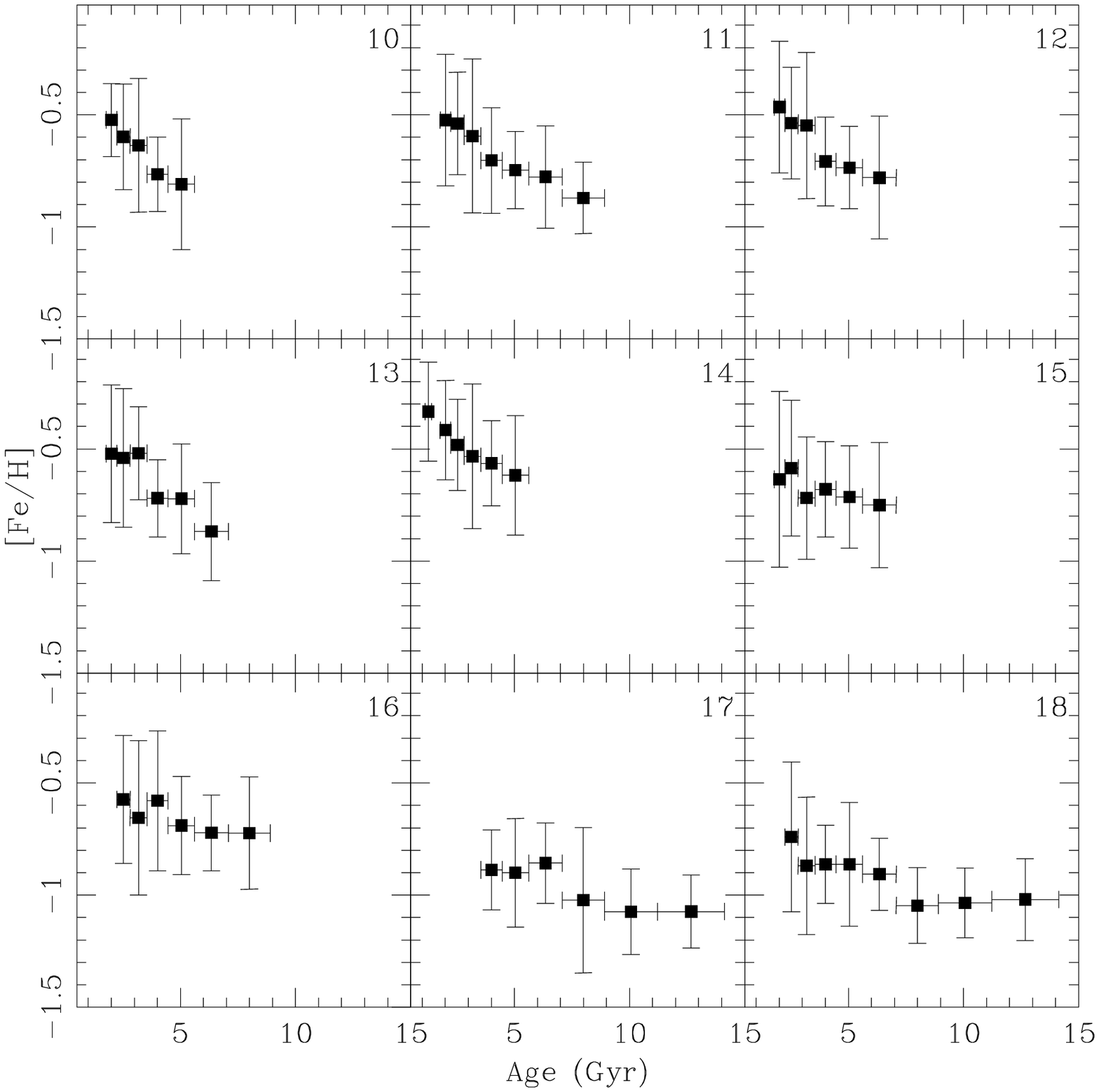}
\caption{continued.}
\label{fig1b}
\end{figure}

\setcounter{figure}{0}
\begin{figure}
\includegraphics[angle=0,width=17cm]{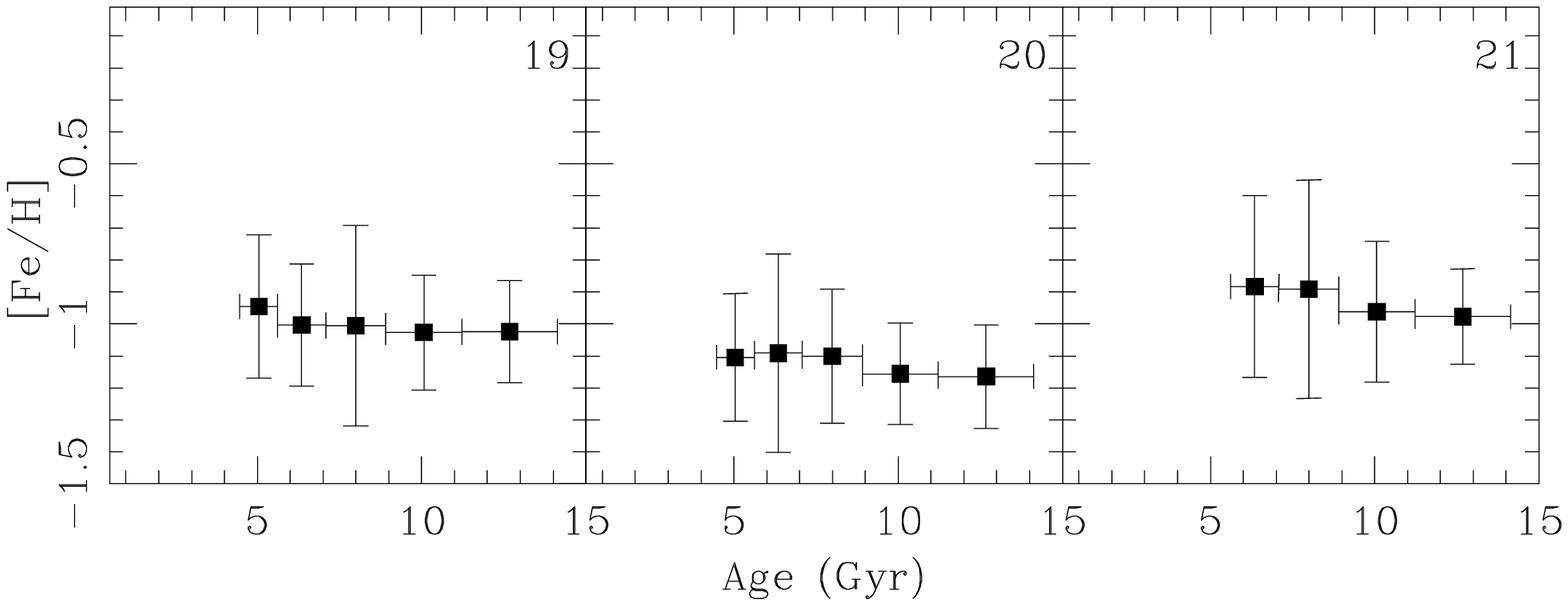}
\caption{continued.}
\label{fig1c}
\end{figure}

\begin{figure}
\includegraphics[angle=0,width=17cm]{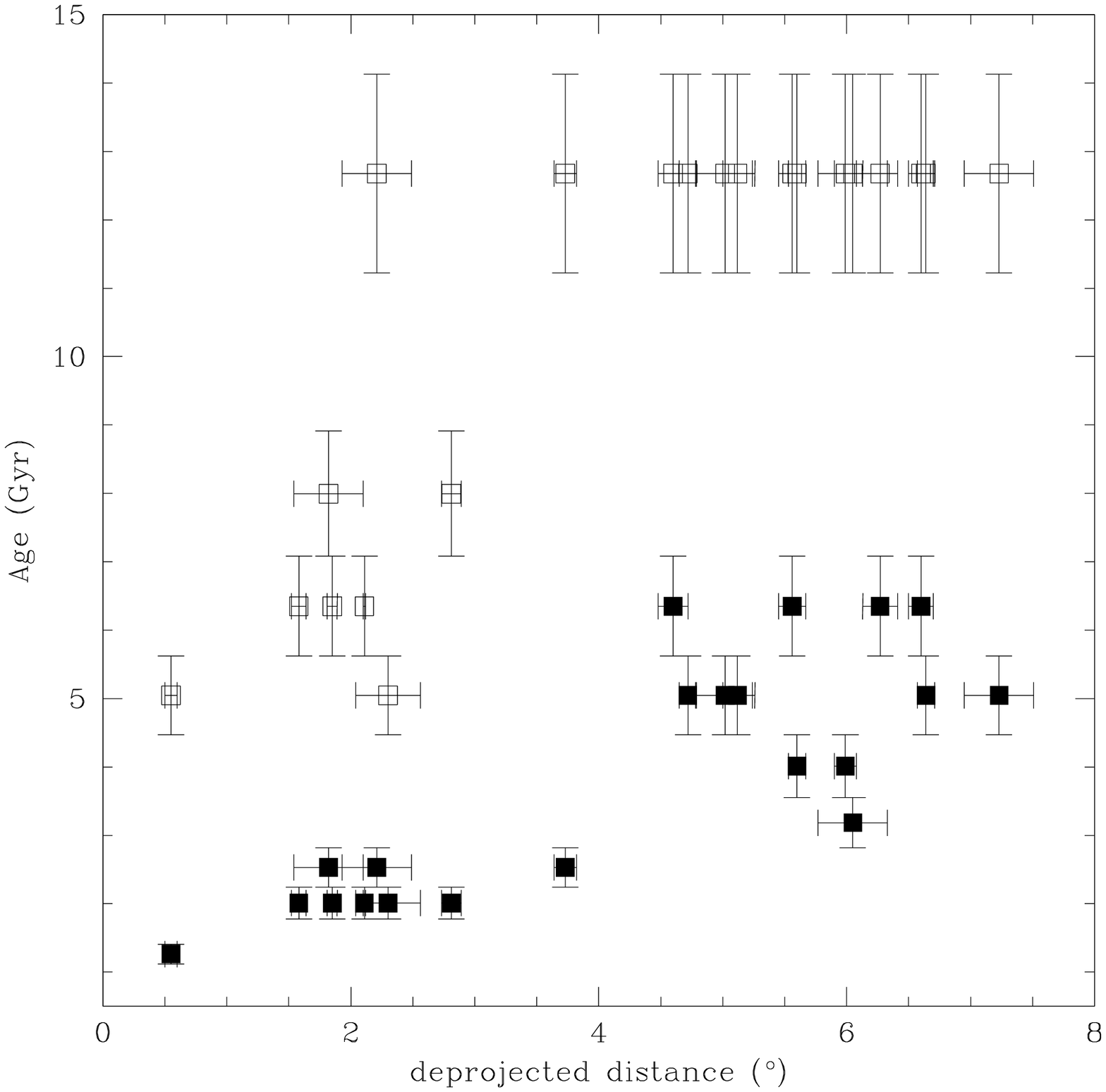}
\caption{Oldest (open box) and youngest (filled box) ages of the AMRs of the 21 studied
LMC fields as a function of the deprojected distance.  Errorbars are as for Fig. 1.}
\label{fig2}
\end{figure}

\begin{figure}
\includegraphics[angle=0,width=17cm]{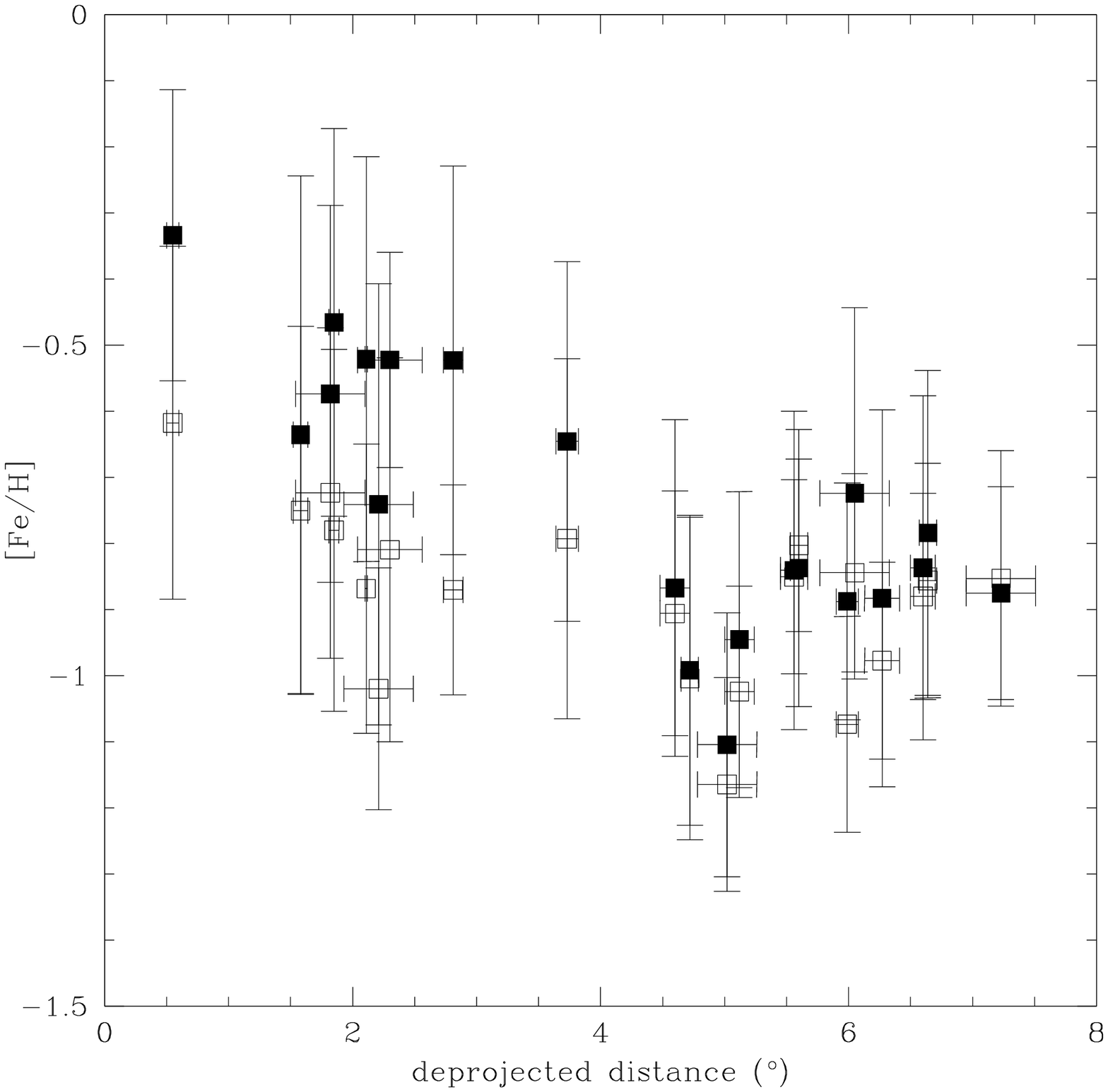}
\caption{Starting (open box) and ending (filled box) metallicities of the AMRs of the 21 studied
LMC fields as a function of the deprojected distance. Errorbars are as for Fig. 1.}
\label{fig3}
\end{figure}

\begin{figure}
\includegraphics[angle=0,width=17cm]{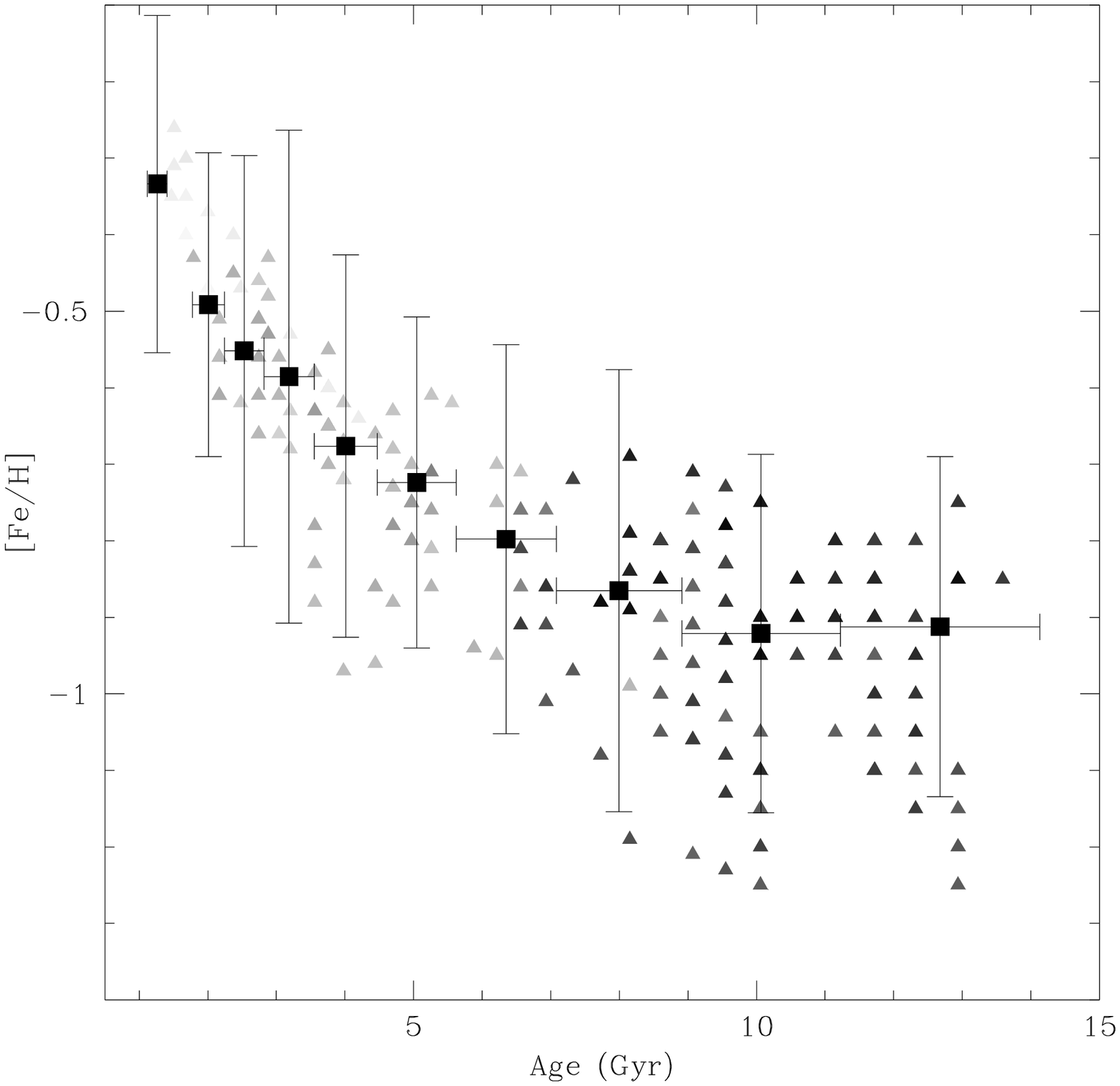}
\caption{Composite AMR for the 21 studied LMC fields. Individual points for the 336 subregions
are also drawn with gray-scale colored triangles: black for the most distant and white for the
nearest fields to the LMC centre. Errorbars are as for Fig. 1.}
\label{fig4}
\end{figure}

\begin{figure}
\includegraphics[angle=0,width=17cm]{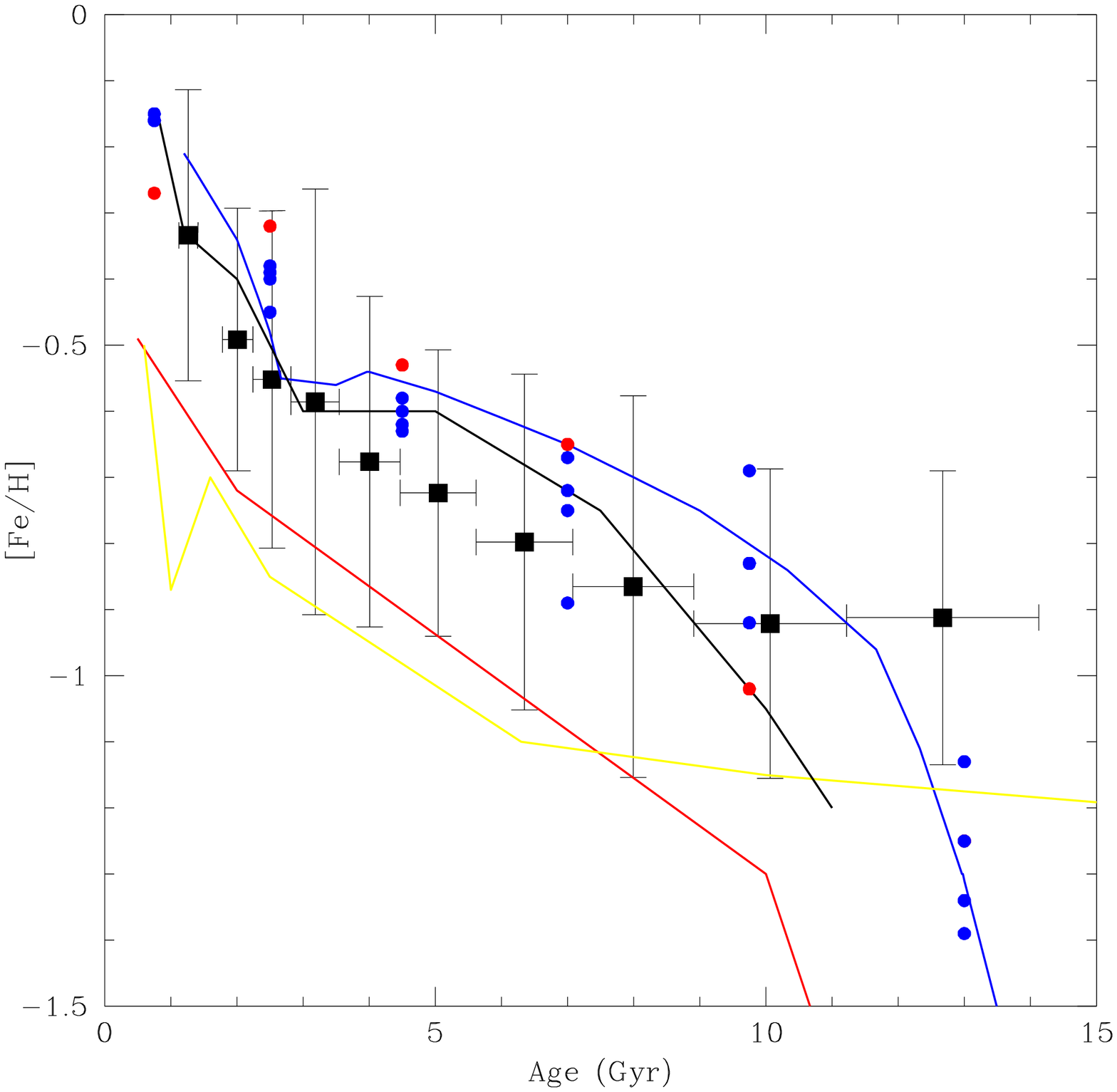}
\caption{Composite AMR for the 21 studied LMC fields as compared with different field AMRs: HZ09 (yellow line),
 \citet{retal11} (black line),  \citet[hereafter PT98)]{pt98} (blue line),
 and \citet{getal98} (red line). 
The red line AMR is based on a closed-box model, while
the blue line relies on a bursting model. We also included with red and blue filled circles the AMRs derived by 
CGAH  for the LMC bar and disk, respectively. Errorbars are as for Fig. 1.}
\label{fig5}
\end{figure}

\begin{figure}
\includegraphics[angle=0,width=17cm]{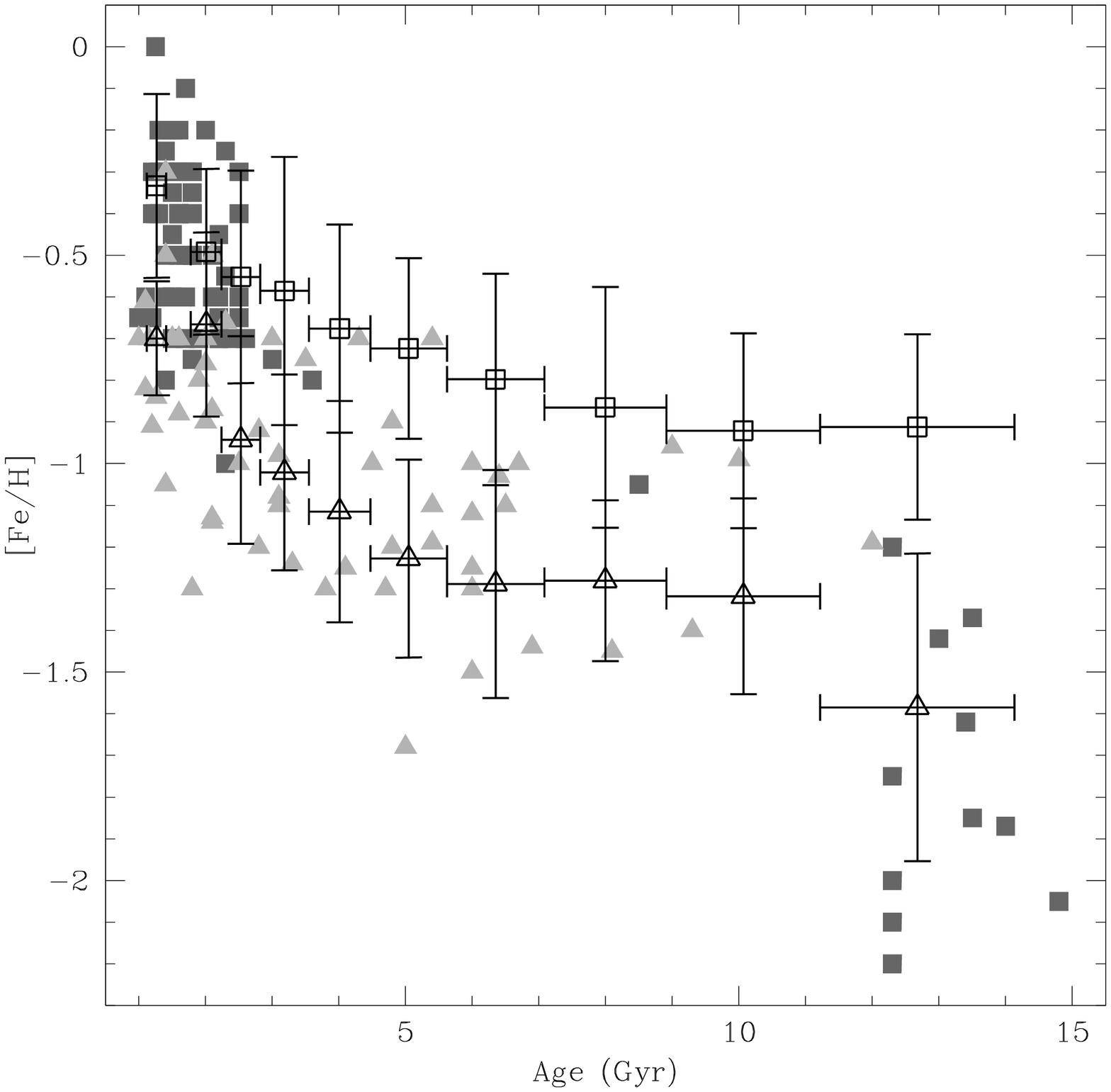}
\caption{Composite field AMRs of the LMC (open boxes) and SMC (open triangles). Their respective cluster AMRs 
are also drawn with filled boxes (LMC) and filled triangles (SMC). Errorbars are as for Fig. 1.}
\label{fig6}
\end{figure}

\begin{deluxetable}{ccccccccccccccccc}
\tabletypesize{\tiny}
\tablecaption{Estimated ages and dispersions (in Gyr) for the representative populations in LMC fields.}
\tablehead{\colhead{Field} & \colhead{A} &    \colhead{B} &    \colhead{C} &   
\colhead{D}  &   \colhead{E}  &   \colhead{F}  &   \colhead{G}  &   \colhead{H}  &   
\colhead{I}  &   \colhead{J}  &   \colhead{K}  &   \colhead{L}  &   \colhead{M}  &   
\colhead{N}  &   \colhead{O}  &    \colhead{P}}

\startdata
 1 &  9.0&  9.0& 11.7&  9.0&  9.0& 11.1& 11.1& 11.7&  9.0& 11.1& 11.1& 11.7&  8.6& 11.1& 11.1& 11.7\\
   &  2.8&  2.8&  3.5&  2.8&  2.8&  3.3&  3.3&  3.5&  2.8&  3.3&  3.3&  3.5&  2.7&  3.3&  3.3&  3.5\\
 2 &  9.5&  9.5& 10.5& 11.1&  8.1& 10.5& 10.5&  8.6& 10.0&  8.1& 10.5& 11.1& 10.5& 10.5&  8.1&  8.1\\
   &  2.9&  2.9&  3.2&  3.3&  2.6&  3.2&  3.2&  2.7&  3.1&  2.6&  3.2&  3.3&  3.2&  3.2&  2.6&  2.6\\
 3 &  8.6& 11.1&  8.6& 11.7&  9.0& 11.1& 11.1&  9.0& 11.7& 11.1&  8.6& 11.7&  8.6&  9.0& 11.7&  9.0\\
   &  2.7&  3.3&  2.7&  3.5&  2.8&  3.3&  3.3&  2.8&  3.5&  3.3&  2.7&  3.5&  2.7&  2.8&  3.5&  2.8\\
 4 & 11.1&  9.5&  9.5&  9.5&  9.5&  7.7& 10.0& 10.5&  9.5& 10.0& 10.0& 10.5& 12.9&  8.1&  8.1&  8.6\\
   &  3.3&  2.9&  2.9&  2.9&  2.9&  2.5&  3.1&  3.2&  2.9&  3.1&  3.1&  3.2&  3.8&  2.6&  2.6&  2.7\\
 5 &  9.5& 11.7& 11.7& 11.7& 11.7& 11.7&  9.0& 11.7& 11.7& 11.7&  9.0&  6.5& 11.7& 12.3&  7.3&  9.5\\
   &  2.9&  3.5&  3.5&  3.5&  3.5&  3.5&  2.8&  3.5&  3.5&  3.5&  2.8&  2.1&  3.5&  3.6&  2.3&  2.9\\
 6 &  9.0&  9.0&  9.0&  9.5& 11.7&  9.0& 11.1&  9.0&  9.0&  9.0& 11.7&  9.0&  9.0& 11.7&  9.0&  9.5\\
   &  2.8&  2.8&  2.8&  2.9&  3.5&  2.8&  3.3&  2.8&  2.8&  2.8&  3.5&  2.8&  2.8&  3.5&  2.8&  2.9\\
 7 & 11.1& 10.0&  4.7& 10.5& 10.5& 10.5&  8.1&  8.1& 10.5& 10.5& 13.5& 10.5& 12.9&  8.1&  8.1& 11.1\\
   &  3.3&  3.1&  1.5&  3.2&  3.2&  3.2&  2.6&  2.6&  3.2&  3.2&  3.9&  3.2&  3.8&  2.6&  2.6&  3.3\\
 8 &  8.6& 11.1&  8.6&  8.6&  9.0& 11.1&  8.6&  9.0&  9.0&  8.6&  8.6&  8.6&  7.3&  6.9&  9.0&  9.0\\
   &  2.7&  3.3&  2.7&  2.7&  2.8&  3.3&  2.7&  2.8&  2.8&  2.7&  2.7&  2.7&  2.3&  2.2&  2.8&  2.8\\
 9 &  6.5&  6.5&  4.9&  5.2&  8.6&  4.9&  3.7&  4.9&  6.5&  8.6&  6.5&  4.9&  9.0&  6.9&  5.2&  6.5\\
   &  2.1&  2.1&  1.6&  1.7&  2.7&  1.6&  1.2&  1.6&  2.1&  2.7&  2.1&  1.6&  2.8&  2.2&  1.7&  2.1\\
 10&  2.7&  2.1&  3.5&  2.7&  3.5&  2.1&  2.1&  2.1&  2.1&  2.1&  2.7&  2.1&  2.7&  2.7&  1.7&  2.1\\
   &  0.8&  0.5&  1.1&  0.8&  1.1&  0.5&  0.5&  0.5&  0.5&  0.5&  0.8&  0.5&  0.8&  0.8&  0.3&  0.5\\
 11&  2.7&  4.7&  3.5&  4.7&  3.5&  2.8&  4.9&  4.9&  4.7&  6.5&  3.7&  4.9&  3.7&  4.9&  3.7&  5.2\\
   &  0.8&  1.5&  1.1&  1.5&  1.1&  0.8&  1.6&  1.6&  1.5&  2.1&  1.2&  1.6&  1.2&  1.6&  1.2&  1.7\\
 12&  2.8&  3.7&  3.5&  2.7&  4.9&  4.9&  3.7&  4.7&  3.7&  3.7&  4.9&  2.8&  3.7&  3.9&  3.9&  5.2\\
   &  0.8&  1.2&  1.1&  0.8&  1.6&  1.6&  1.2&  1.5&  1.2&  1.2&  1.6&  0.8&  1.2&  1.3&  1.3&  1.7\\
 13&  3.0&  2.8&  2.8&  2.8&  3.0&  3.9&  3.7&  3.7&  3.9&  5.2&  3.7&  3.7&  2.3&  3.9&  3.0&  3.7\\
   &  0.9&  0.8&  0.8&  0.8&  0.9&  1.3&  1.2&  1.2&  1.3&  1.7&  1.2&  1.2&  0.6&  1.3&  0.9&  1.2\\
 14&  2.0&  1.6&  1.6&  2.0&  1.5&  1.4&  2.0&  2.3&  1.6&  1.5&  2.4&  4.2&  3.7&  3.2&  3.7&  3.7\\
   &  0.4&  0.3&  0.3&  0.4&  0.2&  0.2&  0.4&  0.6&  0.3&  0.2&  0.7&  1.4&  1.2&  1.0&  1.2&  1.2\\
 15&  3.9&  4.9&  3.7&  3.7&  4.9&  3.9&  3.9&  3.9&  3.7&  3.9&  3.9&  3.2&  3.7&  3.2&  3.0&  2.4\\
   &  1.3&  1.6&  1.2&  1.2&  1.6&  1.3&  1.3&  1.3&  1.2&  1.3&  1.3&  1.0&  1.2&  1.0&  0.9&  0.7\\
 16&  5.5&  4.7&  4.7&  4.9&  6.5&  4.7&  6.2&  6.2&  4.7&  6.2&  6.2&  3.5&  5.2&  6.5&  4.9&  3.7\\
   &  1.8&  1.5&  1.5&  1.6&  2.1&  2.1&  2.0&  2.0&  1.5&  2.0&  2.0&  1.1&  1.7&  2.1&  1.6&  1.2\\
 17&  6.9&  6.9& 11.7& 11.7& 12.3&  9.5&  9.5&  6.5& 10.0& 11.7& 11.7&  9.0& 10.0&  9.5&  9.5&  9.0\\
   &  2.2&  2.2&  3.5&  3.5&  3.6&  2.9&  2.9&  2.1&  3.1&  3.5&  3.5&  2.8&  3.1&  2.9&  2.9&  2.8\\
 18&  4.4&  6.2&  8.1&  8.1&  4.4&  8.1&  8.6& 10.0&  4.4&  4.7&  5.8& 10.0&  3.5&  6.2&  5.8&  4.4\\
   &  1.4&  2.0&  2.6&  2.6&  1.4&  2.6&  2.7&  3.1&  1.4&  1.5&  1.9&  3.1&  1.1&  2.0&  1.9&  1.4\\
 19&  9.5& 11.7&  6.9& 11.7&  9.0& 11.7&  9.0&  9.0&  9.0&  9.0&  9.0&  9.0&  9.0&  6.9& 12.3&  9.5\\
   &  2.9&  3.5&  2.2&  3.5&  2.8&  3.5&  2.8&  2.8&  2.8&  2.8&  2.8&  2.8&  2.8&  2.2&  3.6&  2.9\\
 20& 10.0& 12.9& 10.0& 10.0& 10.0& 10.0& 12.9& 10.0& 10.0& 12.9&  7.7& 12.9& 10.0& 10.0&  7.7&  8.1\\
   &  3.1&  3.8&  3.1&  3.1&  3.1&  3.1&  3.8&  3.1&  3.1&  3.8&  2.5&  3.8&  3.1&  3.1&  2.5&  2.6\\
 21&  8.6& 12.3&  9.5&  9.5& 12.3& 12.3& 12.3& 12.3&  9.5& 12.3&  9.5& 12.3&  9.5&  9.5& 12.3& 10.0\\
   &  2.7&  3.6&  2.9&  2.9&  3.6&  3.6&  3.6&  3.6&  2.9&  3.6&  2.9&  3.6&  2.9&  2.9&  3.6&  3.1\\
\enddata
\end{deluxetable}


\begin{deluxetable}{ccccccccccccccccc}
\tabletypesize{\tiny}
\tablecaption{Estimated metallicities and dispersions (in dex) for the representative populations in LMC fields.}
\tablehead{\colhead{Field} & \colhead{A} &    \colhead{B} &    \colhead{C} &   
\colhead{D}  &   \colhead{E}  &   \colhead{F}  &   \colhead{G}  &   \colhead{H}  &   
\colhead{I}  &   \colhead{J}  &   \colhead{K}  &   \colhead{L}  &   \colhead{M}  &   
\colhead{N}  &   \colhead{O}  &    \colhead{P}}

\startdata
1 & -0.96& -0.86& -0.90& -0.76& -0.91& -1.00& -0.95& -1.00& -0.96& -0.95& -0.95& -0.95& -0.85& -0.95& -0.95& -0.90\\
  &  0.31&  0.31&  0.20&  0.31&  0.31&  0.20&  0.20&  0.20&  0.31&  0.20&  0.20&  0.20&  0.28&  0.20&  0.20&  0.20\\
2 & -0.88& -0.88& -0.90& -0.90& -0.74& -0.90& -0.90& -0.85& -0.80& -0.84& -0.90& -0.85&  --- & -0.90& -0.84& -0.89\\
  &  0.35&  0.35&  0.20&  0.20&  0.26&  0.20&  0.20&  0.28&  0.20&  0.26&  0.20&  0.20&  0.20&  0.20&  0.26&  0.26\\
3 & -0.85& -0.95& -0.90& -0.90& -0.91& -0.95& -0.95& -0.86& -0.95& -0.95& -0.90& -0.90& -0.85& -0.86& -0.90& -0.86\\
  &  0.28&  0.20&  0.28&  0.20&  0.31&  0.20&  0.20&  0.31&  0.20&  0.20&  0.28&  0.20&  0.28&  0.31&  0.20&  0.31\\
4 & -0.90& -0.88& -0.88& -0.93& -0.83& -0.93& -1.00& -1.00& -0.88& -1.00& -0.95& -0.95& -0.90& -0.94& -0.89& -0.90\\
  &  0.20&  0.35&  0.35&  0.35&  0.35&  0.25&  0.20&  0.20&  0.35&  0.20&  0.20&  0.20&  0.20&  0.26&  0.26&  0.28\\
5 & -0.78& -0.90& -0.90& -0.90& -0.90& -0.90& -0.86& -0.90& -0.90& -0.90& -0.86& -0.86& -0.85& -0.85& -0.77& -0.88\\
  &  0.35&  0.20&  0.20&  0.20&  0.20&  0.20&  0.31&  0.20&  0.20&  0.20&  0.31&  0.25&  0.20&  0.20&  0.24&  0.35\\
6 & -0.86& -0.91& -0.86& -0.93& -0.85& -0.96& -0.95& -0.91& -0.91& -0.96& -1.00& -0.91& -0.86& -1.00& -0.91& -1.08\\
  &  0.31&  0.31&  0.31&  0.35&  0.20&  0.31&  0.20&  0.31&  0.31&  0.31&  0.20&  0.31&  0.31&  0.20&  0.31&  0.35\\
7 & -0.85& -0.95& -0.78& -0.95& -0.90& -0.95& -0.89& -0.89& -0.90& -0.95& -0.90& -1.00& -0.80& -0.89& -0.89& -1.00\\
  &  0.20&  0.20&  0.31&  0.20&  0.20&  0.20&  0.26&  0.26&  0.20&  0.20&  0.20&  0.20&  0.20&  0.26&  0.26&  0.20\\
8 & -0.90& -1.00& -0.95& -1.00& -0.96& -1.10& -1.05& -1.06& -1.01& -1.10& -1.05& -1.05& -1.02& -1.06& -1.11& -1.26\\
  &  0.28&  0.20&  0.28&  0.28&  0.31&  0.20&  0.28&  0.31&  0.31&  0.28&  0.28&  0.28&  0.24&  0.24&  0.31&  0.31\\
9 & -0.81& -0.81& -0.80& -0.86& -0.85& -0.85& -0.70& -0.80& -0.81& -0.95& -0.86& -0.80& -0.81& -0.81& -0.76& -0.91\\
  &  0.25&  0.25&  0.30&  0.29&  0.28&  0.30&  0.33&  0.30&  0.25&  0.28&  0.25&  0.30&  0.31&  0.24&  0.29&  0.25\\
10& -0.61& -0.66& -0.88& -0.71& -0.83& -0.61& -0.61& -0.66& -0.66& -0.61& -0.71& -0.61& -0.56& -0.66& -0.48& -0.56\\
  &  0.33&  0.31&  0.34&  0.33&  0.34&  0.34&  0.31&  0.31&  0.31&  0.31&  0.33&  0.31&  0.33&  0.33&  0.28&  0.31\\
11& -0.56& -0.78& -0.68& -0.78& -0.68& -0.58& -0.85& -0.85& -0.83& -0.91& -0.70& -0.80& -0.65& -0.80& -0.70& -0.81\\
  &  0.33&  0.31&  0.34&  0.31&  0.34&  0.34&  0.30&  0.30&  0.31&  0.25&  0.33&  0.30&  0.33&  0.30&  0.33&  0.29\\
12& -0.53& -0.70& -0.68& -0.51& -0.80& -0.85& -0.75& -0.78& -0.70& -0.70& -0.80& -0.53& -0.60& -0.72& -0.72& -0.86\\
  &  0.34&  0.33&  0.34&  0.33&  0.30&  0.30&  0.33&  0.31&  0.33&  0.33&  0.30&  0.34&  0.33&  0.33&  0.33&  0.29\\
13& -0.61& -0.58& -0.58& -0.48& -0.66& -0.77& -0.75& -0.70& -0.77& -0.91& -0.75& -0.70& -0.50& -0.72& -0.66& -0.70\\
  &  0.34&  0.34&  0.34&  0.34&  0.34&  0.33&  0.33&  0.33&  0.33&  0.29&  0.33&  0.33&  0.32&  0.33&  0.34&  0.33\\
14& -0.52& -0.40& -0.45& -0.52& -0.36& -0.40& -0.42& -0.45& -0.35& -0.31& -0.52& -0.69& -0.65& -0.58& -0.65& -0.65\\
  &  0.30&  0.27&  0.27&  0.30&  0.26&  0.25&  0.30&  0.32&  0.27&  0.26&  0.33&  0.32&  0.33&  0.34&  0.33&  0.33\\
15& -0.67& -0.80& -0.70& -0.65& -0.80& -0.77& -0.77& -0.77& -0.70& -0.77& -0.77& -0.73& -0.65& -0.68& -0.71& -0.67\\
  &  0.33&  0.30&  0.33&  0.33&  0.30&  0.33&  0.33&  0.33&  0.33&  0.33&  0.33&  0.34&  0.33&  0.34&  0.34&  0.33\\
16& -0.67& -0.73& -0.68& -0.75& -0.81& -0.73& -0.80& -0.80& -0.73& -0.75& -0.80& -0.63& -0.66& -0.76& -0.75& -0.60\\
  &  0.28&  0.31&  0.31&  0.30&  0.25&  0.31&  0.26&  0.26&  0.31&  0.26&  0.26&  0.34&  0.29&  0.25&  0.30&  0.33\\
17& -0.91& -0.91& -1.15& -1.05& -1.20& -1.18& -1.13& -0.96& -1.25& -1.15& -1.15& -1.06& -1.15& -1.13& -1.13& -1.11\\
  &  0.24&  0.24&  0.20&  0.20&  0.20&  0.35&  0.35&  0.25&  0.20&  0.20&  0.20&  0.31&  0.20&  0.35&  0.35&  0.31\\
18& -0.71& -0.70& -0.74& -0.74& -0.61& -0.74& -0.75& -0.80& -0.71& -0.63& -0.69& -0.80& -0.63& -0.70& -0.69& -0.61\\
  &  0.31&  0.26&  0.26&  0.26&  0.31&  0.26&  0.28&  0.20&  0.31&  0.31&  0.27&  0.20&  0.34&  0.26&  0.27&  0.31\\
19& -1.13& -1.10& -0.96& -0.95& -1.01& -1.15& -1.11& -1.11& -1.06& -1.11& -1.11& -1.11& -0.96& -1.06& -1.15& -1.28\\
  &  0.35&  0.20&  0.24&  0.20&  0.31&  0.20&  0.31&  0.31&  0.31&  0.31&  0.31&  0.31&  0.31&  0.24&  0.20&  0.35\\
20& -1.15& -1.30& -1.30& -1.10& -1.25& -1.30& -1.25& -1.20& -1.20& -1.20& -1.13& -1.15& -1.10& -1.20& -1.13& -1.24\\
  &  0.20&  0.20&  0.20&  0.20&  0.20&  0.20&  0.20&  0.20&  0.20&  0.20&  0.25&  0.20&  0.20&  0.20&  0.25&  0.26\\
21& -0.85& -0.95& -0.93& -0.98& -0.95& -1.00& -1.00& -1.00& -0.93& -1.05& -1.03& -1.00& -0.88& -1.03& -1.10& -1.15\\
  &  0.28&  0.20&  0.35&  0.35&  0.20&  0.20&  0.20&  0.20&  0.35&  0.20&  0.35&  0.20&  0.35&  0.35&  0.20&  0.20\\
\enddata
\end{deluxetable}

\end{document}